\documentclass[numbers,sort&compress,AMA,STIX1COL]{WileyNJD-v2}
\usepackage[T1]{fontenc}
\usepackage[latin9]{luainputenc}
\usepackage{geometry}
\usepackage{color}
\usepackage{float}
\usepackage{textcomp}
\usepackage{amsmath}
\usepackage{amssymb}
\usepackage{graphicx}

\articletype{Research Article}%

\begin{document}

\title{A Comparative Study of Full-Duplex Relaying Schemes for Low Latency Applications}

\author[1]{Fatima Ezzahra Airod}

\author[1]{Houda Chafnaji}

\author[1]{Ahmed Tamtaoui}


\address[]{\orgdiv{Department of Communications systems}, \orgname{INPT}, \orgaddress{\state{Rabat}, \country{Morocco}}}

\corres{Fatima Ezzahra Airod, \\ \email{airod@inpt.ac.ma} \\ Houda Chafnaji, \\ \email{chafnaji@inpt.ac.ma} \\ }


\abstract[Abstract]{Various sectors are likely to carry a set of emerging applications while targeting a reliable communication with low latency transmission. To address this issue, upon a spectrally-efficient transmission, this paper investigates the performance of a one full-dulpex (FD) relay system, and considers for that purpose, two basic relaying schemes, namely the symbol-by-symbol transmission, i.e., amplify-and-forward (AF) and the block-by-block transmission, i.e., selective decode-and-forward (SDF). The conducted analysis presents an exhaustive comparison, covering both schemes, over two different transmission modes, i.e., the non combining mode where the best link, direct or relay link is decoded and the signals combining mode, where direct and relay links are combined at the receiver side. While targeting latency purpose as a necessity, simulations show a refined results of performed comparisons, and reveal that AF relaying scheme is more adapted to combining mode, whereas the SDF relaying scheme is more suitable for non combining mode.}

\keywords{Amplify-and-forward, Selective decode-and-forward, Full-duplex, Low latency applications, Outage probability.}

\maketitle

\section{Introduction}\label{sec1}

An immense amount of data is created every day from different sensors
and peripherals, namely, GPS embedded in vehicles, attached to objects
or worn by people, sensors monitoring the environment, real time video
streams, radars on roads, social network feeds, etc. Such type of
data belongs to real time's domain, where schedulability is one of
the main characteristics of this domain, which means its propensity
to respect the expected time constraints. In fact a real time system
implies a system ability to ensure that investigated processing produces
consistent results, i.e., functionally correct, at the right time.
Therefore, to ensure the radio communication for such applications,
a low latency as well as extreme reliability are required. In this
context, the use of cooperation concept provides spatial and temporal
diversity, and constitutes a good alternative to support advanced
communications with increased channel capacity  \cite{3,4}. 

However, in regards to the end-to-end latency, this requirement
has a significant impact on the system quality and the fluidity of
communications, and it is influenced by different features upon the
transmission, we mention in particular, the propagation delay as well
as the relay delay processing. In fact, depending on the environment
and on the application, we can get rid of some supplementary sources
of delay, as example, for industrial environments such factories,
the distance between two automated robots is not considerable. Hence,
the delay propagation can be neglected, and the only generated delay
in this case, is that related to the relay processing, which depends
mainly on the used relaying technique. 

In general, there are various ways of relay processing in cooperative
networks, among which we distinct mainly two familiar relaying schemes:
amplify-and-forward (AF) and decode-and-forward (DF) \cite{Laneman2}.
In AF scheme, the relay simply amplifies the received signal and forwards
it towards the destination. Thus, in term of the relay processing delay,
the AF scheme, does not include a prominent latency \cite{key-7}.
However, this relaying scheme suffers from noise amplification. In
the DF scheme, the relay first decodes the signal received from the
source, re-encodes and re-transmits it to the destination. This approach
suffers from error propagation when the relay transmits an erroneously
decoded data block. Selective DF (SDF), where the relay only transmits
when it can reliably decode the data packet, has been introduced as
an efficient method to reduce error propagation \cite{Onat}.

In the perspective of a low latency, full-duplex (FD) relaying mode
allows fast device-to-device discovery, and hence, contributes on
the delay reduction. Furthermore, as the capacity improvement is promoted
by the spectral efficiency improvement, the adoption of the 
FD communication at the relay is more advantageous. Even if full-duplex
relaying mode FD generates loop interference from relay input to relay
output, it still practical to use on cooperative relaying system due
to its spectral efficiency \cite{7,8,9}. The FD relay requires the
duplication of radio frequency circuits to transmits and receives
simultaneously in the same time slot and in the same frequency band.
It has been shown that the FD mode still feasible even with the presence
of a significant loop interference \cite{7}, especially with recent
advances noted in antenna technology and signal processing techniques.
In \cite{10}, a novel technique for self-interference cancellation
using antenna cancellation is depicted for FD transmissions. In the
same context, through passive suppression and active self-interference
cancellation mechanisms, an experiment study was proposed in \cite{11}.
Hence, these practical growth incites authors to adopt FD communications
in their research, thus, get rid of spectral inefficiency caused by
half-duplex relaying mode. 
\begin{enumerate}
\item Contributions
\end{enumerate}
Most of previous available works in the literature have investigated
the performance analysis of cooperative networks based SDF and AF
relaying schemes, with the regard to different purposes \cite{12,13,14,15,16,17,18}.
In \cite{12}, considering the FD-AF relaying over Nakagami-$m$ fading
channel, authors cover the performances based on outage probability
and ergodic capacity. The authors in \cite{13,14,15,16,17,18} adopt
a FD cooperative scheme with the direct link between the source and
the destination nodes is non-negligible. Still, in \cite{13,14,15},
to capture the joint benefit of relaying and direct links, at the
destination side, authors have assumed a silence period at the source
that is equal to the processing delay at the relay. The work in \cite{18}, investigates over a Rayleigh fading channel, the optimal mode selection upon a FD-AF system and study therefore, the
individual impact of the residual self-interference (RSI) and the
direct link on the outage performance. However, none
of the cited works have evaluated the relevance in term of latency
impact in the context of latency sensitive applications.

In this paper, we address this issue by conducting a refined comparison
between AF and SDF relaying schemes. Note that each of them adopts a different
block transmission scheme. The pertinence of the direct link effect
is also investigated, through the assumption of two different transmission
modes, i.e., the non combining mode and the signals combining mode. For that
purpose, over the so called Nakagami-$m$ block-fading channel, we
elaborate first the studied transmission schemes communication model,
then we derive their outage probability expressions. Theoretical results
are represented with Monte-carlo simulations and show, on the basis
of a low needed latency, the relevance of each relaying technique,
according to the operating transmission mode. 

The rest of the paper is organized as follows: section \ref{sec:System-Model}
presents the studied system model. The outage probabilities are derived
in section \ref{sec:Amplify-and-Forward-Outage-Proba}. In section
\ref{sec:Results-and-discussion}, numerical performance results are
shown and discussed. The paper is concluded in section \ref{sec:Conclusion}.
\\
\\
\noindent \textcolor{black}{\emph{Notations}}
\begin{itemize}
\item $x$, $\mathbf{x}$, and $\mathbf{X}$ denote, respectively, a scalar
quantity, a column vector, and a matrix.
\item $\mathcal{G}\left(m,\alpha\right)$ represents Gamma distribution
with shape parameter $m$ and rate parameter $\alpha$.
\item $\mathcal{CN}\left(\mu,\sigma\right)$ represents a circularly symmetric
complex Gaussian distribution with mean $\mu$ and variance $\sigma$.
\item $\delta_{m,n}$ is the Kronecker symbol, i.e., $\delta_{m,n}=1$ for
$m=n$ and $\delta_{m,n}=0$ for $m\neq n$.
\item $\left(.\right)^{\star}$\textcolor{black}{,$\left(.\right)^{\top}$,
and $\left(.\right)^{\mathrm{H}}$ are conjugate, the transpose, and
the }Hermitian transpose, respectively\textcolor{black}{. }
\item $\mathbb{C}$ is set of complex number.
\item \textcolor{black}{For $\mathbf{x}\in\mathbb{C}^{N\times1}$, $\mathbf{x}_{f}$
denotes the discrete Fourier transform (DFT) of $\mathbf{x}$, i.e.
$\mathbf{x}_{f}=\mathbf{U}_{N}\mathbf{x}$, with $\mathbf{U}_{N}$
is a unitary $N\times N$ matrix whose $\left(m,n\right)$th element
is $\left(\mathbf{U}_{N}\right)_{m,n}=\frac{1}{\sqrt{N}}e^{-j(2\pi mn/N)}$,
$j=\sqrt{-1}$.}
\item \textcolor{black}{$\left|.\right|$ denotes the }absolute value.
\item $\mathbb{{E}}\left\{ .\right\} $ is used to denote the statistical
expectation.
\item $\mathrm{Pr}\left(X\right)$ is the probability of occurrence of the
event $X$.
\end{itemize}

\section{System Model}\label{sec:System-Model}

This section presents a signal model for one relay cooperative system,
where a FD relay $(\mathrm{R})$, assists the communication
between two end users, representing respectively, the source $(\mathrm{S})$
and the destination $(\mathrm{D})$. In this paper, we assume the
direct link between the source and the destination nodes is non-negligible.
Since relay operates in FD mode, we take into account the RSI
generated from relay's input to relay's output. We consider both,
the non-regenerative and regenerative relaying schemes, namely, amplify-and-forward
and selective decode-and-forward. Hereafter, we introduce first, the
adopted channel model for analysis, then, investigate the system model,
covering both, the AF and the SDF schemes, over two different transmission
modes, i.e., the non-combining mode where the best link, direct or
relay link is decoded and the combining mode, where direct and relay
links are combined at the receiver side.

\subsection{channel model}

The source-destination $\mathrm{S\rightarrow D}$, source-relay $\mathrm{S\rightarrow R}$,
the relay self-interference, and relay-destination $\mathrm{R\rightarrow D}$
channels, are represented by $h_{\mathrm{\mathrm{ab}}},$ with $\mathrm{ab}\epsilon\left\{ \mathrm{SD},\,\mathrm{SR},\,\mathrm{RR},\,\mathrm{RD}\right\} $.
In this paper, we assume that $h_{\mathrm{Z}}$, $Z\epsilon\left\{ \mathrm{SR},\,\mathrm{RD,}\,\mathrm{SD}\right\} $,
are modeled by independent Nakagami-$m$ fading with shape parameter
$m_{\mathrm{Z}}$ and average power $\mathbb{{E}}\left\{ \left|h_{\mathrm{Z}}\right|^{2}\right\} =\sigma_{Z}^{2}$
. Thus, the squared magnitudes $\left|h_{Z}\right|^{2}$ are Gamma
distributed with shape parameter $m_{\mathrm{Z}}$ and rate parameter
$\alpha_{\mathrm{Z}}=\frac{m_{Z}}{\sigma_{\mathrm{Z}}^{2}}$, i.e.,
$\left|h_{\mathrm{Z}}\right|^{2}\sim\mathcal{G}\left(m_{\mathrm{Z}},\alpha_{\mathrm{Z}}\right)$
. The probability density function (PDF) and the cumulative density
function (CDF) of a Gamma random variable $X\sim\mathcal{G}\left(m,\alpha\right)$
are, respectively, given by
\begin{equation}
f_{X}(x)=\frac{\alpha^{m}}{\Gamma(m)}\times x^{m-1}\times e^{-\alpha x},\label{PDF}
\end{equation}
 and
\begin{equation}
F_{X}\left(x\right)=\frac{\gamma(m,\alpha x)}{\Gamma(m)},\qquad x\geq0,\label{CDF}
\end{equation}
where $\Gamma(.)$ denotes the Gamma function and $\gamma(.\,,\,.)$
denotes the lower incomplete Gamma function. 

\subsection{Signal model}

At channel use i, the source node broadcasts its signal $x_{s}(i)$
to both the relay and the destination. Accordingly, the received signal
at the relay and the destination, during channel use i, can be expressed,
respectively, as:

\begin{equation}
y_{\mathrm{R}}(i)=\sqrt{P_{\mathrm{s}}}h_{\mathrm{SR}}x_{\mathrm{s}}(i)+h_{\mathrm{RR}}x_{\mathrm{R}}(i)+n_{\mathrm{R}}(i),\label{eq:combinSR-1}
\end{equation}

\begin{equation}
y_{\mathrm{D}}(i)=\underset{\mathrm{Direct\,signal}}{\underbrace{a\mathrm{_{SD}}x_{\mathrm{s}}(i)}}+\underset{\mathrm{Relayed\,signal}}{\underbrace{a_{\mathrm{RD}}x_{\mathrm{s}}(i-\tau)}}+\eta_{\mathrm{D}}(i),\label{eq:combinDest-2}
\end{equation}
with $a\mathrm{_{SD}=\sqrt{P_{\mathrm{s}}}h_{\mathrm{SD}}}$, $\mathbb{{E}}\left[x_{\mathrm{s}}(i)x_{\mathrm{s}}^{\star}(i')\right]=\delta_{i,i'}$,
$P_{\mathrm{s}}$ denotes the transmit power at the source, $\tau$
is the processing delay at the relay, and $h_{\mathrm{RR}}x_{\mathrm{R}}(i)$
is the RSI after undergoing any
cancellation techniques and practical isolation at the relay \cite{8,19},
and is assumed to be equivalent to a zero mean complex Gaussian random
variable $\sim\mathcal{CN}(0,\sigma_{\mathrm{RR}}^{2})$. $n_{\mathrm{R}}\sim\mathcal{CN}(0,N_{r})$
denotes, a zero-mean complex additive white Gaussian noise at the
relay. Both $a_{\mathrm{RD}}$ and $\eta_{\mathrm{D}}\sim\mathcal{CN}(0,\zeta_{D}^{2})$
depend on the relaying scheme.
\begin{itemize}
\item \textbf{Amplify-and-forward}: With AF scheme, the relay acts as
a repeater which simply amplify the received signal and forwards it
to the destination. Thereby, 
\end{itemize}
\begin{equation}
\begin{cases}
a_{RD}=\beta\sqrt{P_{\mathrm{s}}}h_{\mathrm{RD}}h_{\mathrm{SR}}\\
\eta_{D}(i)=\underset{\mathrm{RSI}}{\underbrace{\beta h_{\mathrm{RD}}h_{\mathrm{RR}}x_{\mathrm{R}}(i-\tau_{\mathrm{AF}})}}+\underset{\mathrm{Noise}}{\underbrace{\beta h_{\mathrm{RD}}n_{\mathrm{R}}(i-\tau_{\mathrm{AF}})+n_{\mathrm{D}}(i)}}
\end{cases}\label{eq:coeffAFDLI}
\end{equation}

where $\beta=\sqrt{\frac{P_{\mathrm{R}}}{P_{\mathrm{s}}|h_{\mathrm{SR}}|^{2}+P_{\mathrm{R}}\sigma_{\mathrm{RR}}^{2}+N_{r}}}$
is the amplification constant factor \textcolor{black}{chosen to satisfy
the total power constraint at the relay} \cite{12}, $P_{\mathrm{R}}$
denotes the transmit power at the relay, $\tau_{AF}$ is the AF relay
processing delay, and $n_{\mathrm{D}}\sim\mathcal{CN}(0,N_{\mathrm{D}})$
denotes a zero-mean complex additive white Gaussian noise at the destination.
\begin{itemize}
\item \textbf{Selective decode-and-forward}: In SDF scheme, the relay
retransmits the received signal only when the $\mathrm{S}\rightarrow\mathrm{R}$
link is not in outage. For this scheme, $\eta_{D}(i)=n_{\mathrm{D}}(i)$
and
\end{itemize}
\begin{equation}
a_\mathrm{RD}=\begin{cases}
\sqrt{P_{\mathrm{R}}}h_{\mathrm{\mathrm{R}D}} & \textrm{if\,\ensuremath{\mathrm{S}\rightarrow\mathrm{R\,}}link\,not\,in\,outage}\\
0 & \textrm{otherwise}
\end{cases}\label{eq:coeffSDFDLI}
\end{equation}

From equation (\ref{eq:combinDest-2}), we see that the destination
node receives the source transmitted signal $x_{\mathrm{s}}$ at different
time instances due to the processing delay $\tau$ at the relay. In
this work, we consider two transmission modes: Non combining (NC)
mode where the receiver is synchronized with the strongest link, direct
or relay link, and Signals combining (SC) mode where both
direct and relay links are combined at the receiver side.

\subsubsection{Non Combining (NC) mode }

In this mode, the destination will try to decode the strongest link
while, the second one will be considered as interference. Therefore,
the system capacity of the NC mode for AF and SDF is expressed respectively,
as:

\begin{equation}
C_{NC}^{AF}=\mathrm{log_{2}\left(1\text{+}\gamma_{AF}\right)}\label{eq:af_dli-1}
\end{equation}
\begin{equation}
C_{NC}^{SDF}=\mathrm{log_{2}\left(1\text{+}\gamma_{SDF}\right),}\label{eq:sdf_dli}
\end{equation}
with $\mathrm{\gamma_{AF}}=\frac{|a\mathrm{_{XD}}|^{2}}{|a_{\mathrm{\overline{X}D}}|^{2}+\beta^{2}|h_{\mathrm{RD}}|^{2}(P_{\mathrm{R}}\sigma_{\mathrm{RR}}^{2}+N_{\mathrm{r}})+N_{D}}$
and $\mathrm{\gamma_{SDF}}=\frac{P_{\mathrm{R}}|h_{\mathrm{XD}}|^{2}}{P_{\mathrm{\mathrm{s}}}|h_{\mathrm{\overline{X}D}}|^{2}+N_{D}}$,
are respectively, the AF and SDF signal-to-interference and noise
ratio (SINR), where $\mathrm{X}\rightarrow\mathrm{D}$ is the best
link and $\mathrm{\overline{X}}\rightarrow\mathrm{D}$ is the worst
link considered as interference.

\subsubsection{Signals Combining (SC) mode }

In SC mode both relay and direct signals are combined at the destination
side. Therefore, in order to alleviate the inter-symbol interference
(ISI) caused by the delayed signal, equalization is performed at the
destination. For that purpose, we propose a cyclic-prefix (CP) transmission
at the source side in order to perform frequency-domain equalization
(FDE) at the destination node. Depending on the processing protocol
at the relay, AF or SDF, the destination performs signal-based FDE
or block-based FDE. In the following, we assume all channel gains
remain constant during $N+\tau_{\mathrm{CP}}$ channel uses\footnote{$N+\tau_{\mathrm{CP}}$ is less or equal to the channel coherence
time $T_{c}$. For simplicity, we assume that all links have the same
$T_{c}$}, where $\tau_{\mathrm{CP}}$ is the CP length ($\tau_{\mathrm{CP}}\geq\mathit{\tau}$). 

At the destination side, after the CP removal, the received signal,
at channel use $i$, can be expressed as, 

\begin{equation}
y_{\mathrm{D}}(i)=\underset{\mathrm{Direct\,+\,Relayed\,signal}}{\underbrace{\sqrt{P_{\mathrm{s}}}h_{\mathrm{SD}}x_{\mathrm{s}}(i)+a_{\textrm{RD}}x_{\mathrm{s}}\left[(i-\tau)\,\mathrm{mod}\,N\right]}}+\eta_{\mathrm{D}}(i),\label{eq:combinDest-1}
\end{equation}

Equation (\ref{eq:combinDest-1}) can be written in vector form to
jointly take into account the $N+\tau$ received signal as: 

\begin{equation}
\mathbf{y}_{D}=\boldsymbol{{\mathcal{H}}}\mathbf{x}_{\mathrm{s}}+\boldsymbol{\mathbf{\eta_{D}}},
\end{equation}

where $\mathbf{y}_{D}=\left[y_{D}\left(0\right),...,y_{D}\left(N-1\right)\right]^{\top}\in\mathbb{C}^{N\times1}$,
$\mathbf{x}_{\mathrm{s}}=\left[x_{\mathrm{s}}\left(0\right),...,x_{\mathrm{s}}\left(N-1\right)\right]^{\top}\in\mathbb{C}^{N\times1}$,
$\mathbf{\boldsymbol{\eta}_{D}}=\left[\eta_{\mathrm{D}}\left(0\right),...,\eta_{\mathrm{D}}\left(N-1\right)\right]^{\top}\in\mathbb{C}^{N\times1}$
, and $\boldsymbol{{\mathcal{H}}}\in\mathbb{C}^{N\times N}$ is a
circulant matrix whose first column matrix is $\left[a\mathrm{_{SD}},\boldsymbol{{0}}_{1\times\tau-2},a_{\textrm{RD}},\boldsymbol{{0}}_{1\times N-\tau}\right]^{\top}$.
Note that the circulant matrix $\boldsymbol{{\mathcal{H}}}$ , can
be decomposed as, $\boldsymbol{{\mathcal{H}}}=\mathbf{U}_{N}^{H}\boldsymbol{{\Lambda}}\mathbf{U}_{N}$.
With $\boldsymbol{{\Lambda}}$ is a diagonal matrix whose $\left(i,i\right)$-th
element is $\lambda_{i}=a\mathrm{_{SD}}+a_{\textrm{RD}}e^{-j\left(2\pi i\frac{\tau}{N}\right)}.$
Therefore, the signal $\mathbf{y}_{\mathrm{D}}$ can be represented
in the frequency domain as,

\begin{equation}
\mathbf{y}_{\mathrm{D}_{f}}=\boldsymbol{{\Lambda}}\mathbf{x}_{\mathrm{s}_{f}}+\mathbf{n}_{f}.
\end{equation}

At the destination side, the system capacity is given by, 

\begin{equation}
C_{\mathrm{SC}}=\left(\frac{N}{N+\tau}I(\mathbf{x}_{\mathrm{s}_{f}},\mathbf{y}_{\mathrm{D}_{f}})\right),\label{eq:14-2}
\end{equation}
 where the factor $\frac{N}{N+\tau}$ means that the transmission
of $N$ useful bits occupies $N+\tau$ channel uses and $I$ represents
the overall system average mutual information, and is given by, $I(\mathbf{x}_{\mathrm{s}_{f}},\mathbf{y}_{\mathrm{D}_{f}})=\frac{1}{N}{\displaystyle \sum_{i=0}^{N-1}}\mathrm{log_{2}}(1+\gamma_{i})$,
where $\gamma_{i}=\frac{\lambda_{i}\lambda_{i}^{H}}{\zeta_{D}^{2}}=\rho+2\left|\mu\right|cos\left(2\pi i\frac{\tau}{N}+\theta\right)$,
with $\rho=\frac{|a_{\mathrm{\mathrm{S}D}}|^{2}+|a_{\mathrm{RD}}|^{2}}{\zeta_{D}^{2}}$,
$\mu=\frac{a_{\mathrm{\mathrm{S}D}}a_{\mathrm{RD}}}{\zeta_{D}^{2}}$,
and $\theta=angle\left(h\mathrm{_{SD}},h_{\mathrm{RD}}^{*}\right)$.

The system mutual information $I(\mathbf{x}_{\mathrm{s}_{f}},\mathbf{y}_{\mathrm{D}_{f}})$
can be manipulated as below,

\begin{align}
I(\mathbf{x}_{\mathrm{s}_{f}},\mathbf{y}_{\mathrm{D}_{f}}) & =\frac{1}{N}\sum_{i=0}^{N-1}\mathrm{log_{2}}(1+\gamma_{i})\nonumber \\
 & =\frac{1}{N}\sum_{i=0}^{N-1}\mathrm{log_{2}}\left[(1+\rho)\left(1+\frac{2\left|\mu\right|cos\left(2\pi i\frac{\tau}{N}+\theta\right)}{1+\rho}\right)\right]\label{eq:Mutual_Info}\\
 & =\mathrm{log_{2}}(1+\rho)+\frac{1}{N}\sum_{i=0}^{N-1}\mathrm{log_{2}}\left(1+\frac{2\left|\mu\right|cos\left(2\pi i\frac{\tau}{N}+\theta\right)}{1+\rho}\right).\nonumber 
\end{align}
According to the arithmetic-geometric mean inequality, $\rho\geq2\left|\mu\right|$,
we have $1+\rho>2\left|\mu\right|cos\left(2\pi i\frac{\tau}{N}+\theta\right).$
Thus, by using the first order Taylor expansion, we have $\mathrm{ln}\left(1+\frac{2\left|\mu\right|cos\left(2\pi i\frac{\tau}{N}+\theta\right)}{1+\rho}\right)\thickapprox\frac{2\left|\mu\right|cos\left(2\pi i\frac{\tau}{N}+\theta\right)}{1+\rho}.$
Noting that ${\displaystyle \sum_{i=0}^{N-1}}cos\left(2\pi i\frac{\tau}{N}+\theta\right)=0$,
the mutual information, in (\ref{eq:Mutual_Info}), can be approximated
as,

\begin{align}
I(\mathbf{x}_{\mathrm{s}_{f}},\mathbf{y}_{\mathrm{D}_{f}}) & \thickapprox\mathrm{log_{2}}(1+\rho)\label{eq:approx_snr}
\end{align}

\begin{figure*}[tbh]
\begin{centering}
\includegraphics[scale=0.5]{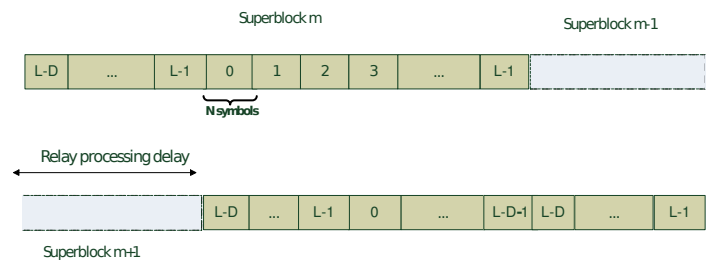}
\par\end{centering}
\caption{\label{fig:SDF-Block-transmission}SDF Block transmission scheme}

\end{figure*}

\begin{itemize}
\item \textbf{Amplify-and-Forward}: AF is classified as memoryless scheme
in which the relay processes the received signal in a symbol-by-symbol
manner. Therefore, the processing delay $\tau_{AF}$ is in term of
channel uses and thus, the equalization, at the destination side,
is a signal-based equalization.
\item \textbf{Selective Decode and Forward}: Unlike AF, SDF is a memory
scheme where the entire received block need to be decoded, before
deciding to retransmit or not the re-encoded block through $\mathrm{R}\rightarrow\mathrm{D}$
link. This results in a block-based processing delay $\tau_{\mathrm{SDF}}$.
Thereby, as depicted in Fig.\ref{fig:SDF-Block-transmission}, to
deal with inter-block interferences, the communication takes place
assuming one superblock transmission of $L$ blocks, each gathering
$N$ symbols and the SDF CP prefix is constructed using D blocks of
$N$, i.e., $\tau_{\mathrm{SDF}}=D\times N$. 
\end{itemize}
\section{Outage Probability}\label{sec:Amplify-and-Forward-Outage-Proba}

In this section, we present the outage analysis of different schemes investigated in section \ref{sec:System-Model}. The system outage occurs when the received SINR at the destination side is below a target SNR threshold, whether
for SC mode, where both relay and direct signals are combined at the
destination side, or NC mode, where the destination will try to decode
the strongest link while, the second one will be considered as interference.
Note that, in this work, the packet re-transmission is not considered. 
\\Hereafter, we derive first for each mode, i.e. NC and SC, the overall full-duplex outage probability
of the AF scheme as well as that of the SDF \cite{14}. For the purpose
of investigating the analysis, let's first introduce the instantaneous
SINRs for each link. 

The received SINR of the $\mathrm{S}\rightarrow\mathrm{D}$, the $\mathrm{S}\rightarrow\mathrm{R}$
and the $\mathrm{R}\rightarrow\mathrm{D}$ links are denoted, respectively,
as,

\begin{equation}
\begin{aligned}\gamma_{\mathsf{S\mathrm{D}}} & =\frac{P_{\mathrm{s}}|h_{\mathrm{SD}}|^{2}}{N_{\mathrm{D}}},\,\gamma_{\mathrm{SR}}=\frac{P_{\mathrm{s}}|h_{\mathrm{\mathrm{SR}}}|^{2}}{P_{\mathrm{R}}\sigma_{\mathrm{RR}}^{2}+N_{\mathrm{r}}},\,\gamma_{\mathrm{RD}}=\frac{P_{\mathrm{R}}|h_{\mathrm{RD}}|^{2}}{N_{\mathrm{D}}}.\end{aligned}
\label{eq:linkSNRNC}
\end{equation}

Note that $\gamma_{\mathrm{SR}}$, $\gamma_{\mathrm{SD}}$ and $\gamma_{\mathrm{RD}}$
are the result of a Gamma random variable scaled by a constant. Therefore,
$\gamma_{\mathrm{SR}}$, $\gamma_{\mathrm{SD}}$ and $\gamma_{\mathrm{RD}}$
are Gamma distributed with shape parameter $m_{\mathrm{Z}}$ and rate
parameter $\tilde{\alpha}_{\mathrm{Z}}$, where $\tilde{\alpha}_{\mathrm{SD}}=\frac{N_{\mathrm{D}}}{P_{\mathrm{S}}}\alpha_{\mathrm{SD}}$,
$\tilde{\alpha}_\mathrm{{SR}}=\frac{P_{\mathrm{R}}\sigma_{\mathrm{RR}}^{2}+N_{\mathrm{r}}}{P_{\mathrm{S}}}\alpha_{\mathrm{SR}}$
and $\tilde{\alpha}_{\mathrm{RD}}=\frac{N_{r}}{P_{\mathrm{R}}}\alpha_{\mathrm{RD}}$.

Herein, the outage probability is denoted $P_{\mathrm{out}}$ and
expressed as:
\begin{equation}
\begin{aligned}P_{\mathrm{out}}= & \mathrm{Pr}(\gamma<\eta)\\
= & F_{\gamma}\left(\eta\right),
\end{aligned}
\label{eq:outageNCAF}
\end{equation}
where $\eta=2^{R}-1$, with $R$ is the bit rate per channel use,
and $F_{\gamma}\left(.\right)$ is the CDF of $\gamma$. 

\subsection{Non combining mode }
Herein, for the rest of NC mode analysis, we consider $\eta_{AF}=\eta_{SDF}=\eta=2^{R}-1$.
\begin{itemize}
\item \textbf{Amplify-and-forward}
\end{itemize}
For AF relaying scheme, using (\ref{eq:linkSNRNC}), we extract the
corresponding end-to-end SINR as, 
\begin{equation}
\mathrm{\gamma_{AF}}=\begin{cases}
\frac{\gamma_{\mathrm{SR}}\frac{\gamma_{\mathrm{RD}}}{\gamma_{\mathrm{SD}+1}}}{\gamma_{\mathrm{SR}}+\frac{\gamma_{\mathrm{RD}}}{\gamma_{\mathrm{SD}+1}}+1}. & if\,\gamma_{\mathrm{RD}}>\gamma_{\mathrm{SD}}\\
\frac{\gamma_{\mathrm{SD}}}{\gamma_{\mathrm{RD}+1}} & otherwise
\end{cases}
\end{equation}
 Herein, for the case where the $\mathrm{R}\rightarrow\mathrm{D}$
link is stronger than the $\mathrm{S\rightarrow D}$, after some manipulations,
the equation (\ref{eq:outageNCAF}), tends to an integral form which doesn't
generate a closed form expression, and can be evaluated numerically
using matlab software. Otherwise, using \cite[Eq 12]{14}, the outage
probability can be derived as,

\begin{equation}
F_{\mathrm{\gamma_{AF}}}(\eta)=\frac{\gamma(m_{\mathrm{SD}},\tilde{\alpha}_{\mathrm{SD}}\eta)}{\Gamma(m_{\mathrm{SD}})}+B\sum_{k=0}^{m_{\mathrm{RD}}-1}c^{-d}\tilde{\alpha}_{\mathrm{\mathrm{RD}}}^{k}W_{a,b}[c],\label{eq:SDF_DLISRD-1}
\end{equation}
where $W_{a,b}[c]$ denotes, the Whittaker function, $a=\frac{m_{\mathrm{SD}}-k-1}{2}$,
$b=\frac{-m_{\mathrm{SD}}-k}{2}$, $c=\tilde{\alpha}_{\mathrm{SD}}\eta+\tilde{\alpha}_{\mathrm{\mathrm{RD}}}$,
$d=\frac{m_{\mathrm{SD}}+k+1}{2}$, and $B=\frac{e^{-\left(\frac{1}{2}\left(\tilde{\alpha}_{\mathrm{SD}}\eta-\tilde{\alpha}_{\mathrm{\mathrm{RD}}}\right)\right)}}{\Gamma(m_{\mathrm{SD}})}\left(\tilde{\alpha}_{\mathrm{SD}}\eta\right)^{m_{\mathrm{SD}}}$.
\begin{itemize}
\item \textbf{Selective Decode and Forward}
\end{itemize}
The SDF relay system outage probability, is generally, defined as:

\begin{equation}
\begin{aligned}\mathcal{P}_{\mathrm{out}} & =P_{\mathrm{out}}^{\mathrm{S\rightarrow D}}P_{\mathrm{out}}^{\mathrm{S\rightarrow R}}+\mathit{P_{\mathrm{out}}^{\mathcal{\mathrm{X}}\mathrm{D}}}\left(1-P_{\mathrm{out}}^{\mathrm{S\rightarrow R}}\right)\end{aligned}
,\label{eq:sdfout}
\end{equation}
where $P_{\mathrm{out}}^{\mathrm{S\rightarrow D}}$ and $P_{\mathrm{out}}^{\mathrm{S\rightarrow R}}$
denote respectively, the outage probability of $\mathrm{S\rightarrow D}$
link and $\mathrm{S\rightarrow R}$ link, and can be expressed as
in \cite{14},
\begin{eqnarray}
P_{\mathrm{out}}^{\mathrm{S\rightarrow D}}= & \frac{\gamma(m_{\mathrm{SD}},\tilde{\alpha}_{\mathrm{SD}}\eta)}{\Gamma(m_{\mathrm{SD}})}\label{eq:SD_cdf}\\
P_{\mathrm{out}}^{\mathrm{S\rightarrow R}}= & \frac{\gamma(m_{\mathrm{SR}},\tilde{\alpha}_{\mathrm{SR}}\eta)}{\Gamma(m_{\mathrm{SR}})}\label{eq:SR_SD}
\end{eqnarray}
 $\mathit{P_{\mathrm{out}}^{\mathcal{\mathrm{X}}\mathrm{D}}}$ denotes
the outage probability of the best link $\mathrm{X\rightarrow D}$,
i.e., $\mathrm{R\rightarrow D}$ or $\mathrm{S\rightarrow D}$, when
the relay correctly decodes the received signal, and it can be derived
as follows, 

\begin{align}
\mathit{P_{\mathrm{out}}^{\mathrm{X}\mathrm{D}}}= & \mathrm{Pr}\left(\gamma_{\mathrm{SDF}}<\eta\right)\nonumber \\
= & F_{\gamma_{\mathrm{SDF}}}(\eta)\label{eq:dlsdf}
\end{align}
and it is given by the following expression \cite[Eq. 12]{14}: 

\begin{equation}
F_{\gamma_{\mathrm{SDF}}}(\eta)=\frac{\gamma(m_{\mathrm{XD}},\tilde{\alpha}_{\mathrm{XD}}\eta)}{\Gamma(m_{\mathrm{XD}})}+B\sum_{k=0}^{m_{\mathrm{\overline{X}D}}-1}c^{-d}\tilde{\alpha}_{\mathrm{\mathrm{\overline{X}D}}}^{k}W_{a,b}[c],\label{eq:SDF_DLISRD}
\end{equation}
where $W_{a,b}[c]$ denotes, the Whittaker function, $a=\frac{m_{\mathrm{XD}}-k-1}{2}$,
$b=\frac{-m_{\mathrm{XD}}-k}{2}$, $c=\tilde{\alpha}_{\mathrm{XD}}\eta+\tilde{\alpha}_{\mathrm{\mathrm{\overline{X}D}}}$,
$d=\frac{m_{\mathrm{XD}}+k+1}{2}$, and $B=\frac{e^{-\left(\frac{1}{2}\left(\tilde{\alpha}_{\mathrm{XD}}\eta-\tilde{\alpha}_{\mathrm{\mathrm{\overline{X}D}}}\right)\right)}}{\Gamma(m_{\mathrm{XD}})}\left(\tilde{\alpha}_{\mathrm{XD}}\eta\right)^{m_{\mathrm{XD}}}$.

Finally, by substituting (\ref{eq:SD_cdf}), (\ref{eq:SR_SD}), and
(\ref{eq:SDF_DLISRD}) into (\ref{eq:sdfout}), we get the closed
form expression of NC SDF outage probability.

\subsection{Signals Combining mode }
\begin{itemize}
\item \textbf{\label{Amplify-and-forward_SC_out_exp}Amplify-and-forward}
\end{itemize}
The outage probability of FD AF combining system is derived as follows:

\begin{eqnarray}
P_{\mathrm{out}} & = & \mathrm{Pr}\left(\frac{N}{N+\tau_{AF}}\mathrm{log_{2}}(1+\rho)<r\right)\nonumber \\
 & = & F_{\rho}\left(\eta_{AF}\right),\label{eq:15}
\end{eqnarray}
where $\eta_{AF}=2^{R\left(\frac{N+\tau_{AF}}{N}\right)}-1$. By substituting
(\ref{eq:linkSNRNC}) into $\rho$, we get,

\begin{equation}
\begin{aligned}\rho= & \frac{\gamma_{\mathrm{SR}}\gamma_{\mathrm{RD}}+\gamma_{\mathrm{SD}}\gamma_{\mathrm{SR}}+\gamma_{\mathrm{SD}}}{1+\gamma_{\mathrm{SR}}+\gamma_{\mathrm{RD}}}.\end{aligned}
\end{equation}
Thus, the CDF of $\rho$ can be derived as,

\begin{align}
F_{\rho}\left(x\right) & =\mathrm{Pr}(\rho<x)=\mathrm{Pr}\left(\gamma_{\mathrm{SD}}\leq\underset{\mathrm{\Delta}}{\underbrace{\frac{(1+\gamma_{\mathrm{SR}}+\gamma_{\mathrm{RD}})x-\gamma_{\mathrm{SR}}\gamma_{\mathrm{RD}}}{1+\gamma_{\mathrm{SR}}}}}\right)\nonumber \\
 & =\iint_{\Delta\geq0}\frac{\gamma(m_{\mathrm{SD}},\tilde{\alpha}_{\mathrm{SD}}\Delta)}{\Gamma(m_{\mathrm{SD}})}\times f_{\gamma_{\mathrm{SR}}}(y)\times f_{\gamma_{\mathrm{RD}}}(z)dydz.\label{eq:19}
\end{align}
where $\gamma(l,x)$ presents the lower incomplete Gamma function
\cite{20}.

Hereafter, to solve this double integral, we need to decompose the integration into two steps. Therefore, in order to proceed, let's denote $g(\gamma_{\mathrm{SR}},z)=\frac{\gamma(m_{\mathrm{SD}},\tilde{\alpha}_{\mathrm{SD}}\Delta)}{\Gamma(m_{\mathrm{SD}})}\times f_{\gamma_{\mathrm{RD}}}(z)$.
First, while treating $\gamma_{\mathrm{SR}}$ as constant, we have to integrate $g(\gamma_{\mathrm{SR}},z)$ with
respect to the limits $\begin{cases}
z_{1}=\left(\frac{1+\gamma_{\mathrm{SR}}}{\gamma_{\mathrm{SR}}-\eta_{AF}}\right)\eta_{AF}\,\textrm{and}\,z_{2}=\infty, & \textrm{if}\,\gamma_{\mathrm{SR}}\leq\eta_{AF}\\
z_{1}=0\,\textrm{and}\,z_{2}=\left(\frac{1+\gamma_{\mathrm{SR}}}{\gamma_{\mathrm{SR}}-\eta_{AF}}\right)\eta_{AF}, & \textrm{if}\,\gamma_{\mathrm{SR}}>\eta_{AF}
\end{cases}$. Using the serie form, i.e. $\gamma(l,x)=(l-1)!\left[1-e^{-x}{\displaystyle \sum_{a=0}^{l-1}}\frac{x^{a}}{a!}\right]$
\cite[8.352.6]{20} and the polynomial expansion, i.e. $(x+1)^{a}={\displaystyle \sum_{b=0}^{a}}\frac{a!}{b!(a-b)!}x^{b}$,
we get therefore, the first integral resolution, i.e., $G\left(\eta_{AF}\right)$
as represented in (\ref{eq:System_CDF-1}). 
\begin{algorithm*}[tbh]
\begin{equation}
G\left(\eta_{AF}\right)=\begin{cases}
1-{\displaystyle \sum_{a=0}^{m_{\mathrm{SD}}-1}}{\displaystyle \sum_{b=0}^{a}}\frac{\left(\frac{\eta_{AF}-\gamma_{\mathrm{SR}}}{\eta_{AF}\left(1+\gamma_{\mathrm{SR}}\right)}\right)^{b}\times\left(\tilde{\alpha}_{\mathrm{SD}}\eta_{AF}\right)^{a}}{b!(a-b)!}\times\frac{\Gamma\left(m_{\mathrm{RD}}+b\right)}{\left(\tilde{\alpha}_{\mathrm{RD}}+\left(\frac{\eta_{AF}-\gamma_{\mathrm{SR}}}{1+\gamma_{\mathrm{SR}}}\right)\tilde{\alpha}_{\mathrm{SD}}\right)^{(m_{\mathrm{RD}}+b)}}\times\frac{e^{(-\tilde{\alpha}_{\mathrm{SD}}\eta_{AF})}\times\tilde{\alpha}_{\mathrm{RD}}^{m_{\mathrm{RD}}}}{\Gamma(m_{\mathrm{RD}})}, & \gamma_{\mathrm{SR}}\leq\eta_{AF}\\
\frac{\gamma\left(m_{RD},\tilde{\alpha}_{\mathrm{RD}}\left(\frac{1+\gamma_{\mathrm{SR}}}{\gamma_{\mathrm{SR}}-\eta_{AF}}\right)\eta_{AF}\right)}{\Gamma(m_{\mathrm{RD}})}-{\displaystyle \sum_{a=0}^{m_{\mathrm{SD}}-1}}{\displaystyle \sum_{b=0}^{a}}\frac{\left(\frac{\eta_{AF}-\gamma_{\mathrm{SR}}}{\eta_{AF}\left(1+\gamma_{\mathrm{SR}}\right)}\right)^{b}\times\left(\tilde{\alpha}_{\mathrm{SD}}\eta_{AF}\right)^{a}}{b!(a-b)!}\times\frac{\gamma\left(m_{\mathrm{RD}}+b,\tilde{\alpha}_{\mathrm{RD}}\left(\frac{1+\gamma_{\mathrm{SR}}}{\gamma_{\mathrm{SR}}-\eta_{AF}}\right)\eta_{AF}-\tilde{\alpha}_{\mathrm{SD}}\eta_{AF}\right)}{\left(\tilde{\alpha}_{\mathrm{RD}}+\left(\frac{\eta_{AF}-\gamma_{\mathrm{SR}}}{1+\gamma_{\mathrm{SR}}}\right)\tilde{\alpha}_{\mathrm{SD}}\right)^{(m_{\mathrm{RD}}+b)}}\\
\times\frac{e^{\left(-\tilde{\alpha}_{\mathrm{SD}}\eta_{AF}\right)}\times\tilde{\alpha}_{\mathrm{RD}}^{m_{\mathrm{RD}}}}{\Gamma(m_{\mathrm{RD}})}, & \gamma_{\mathrm{SR}}>\eta_{AF}
\end{cases}\label{eq:System_CDF-1}
\end{equation}
\end{algorithm*}
Now, the resulting expression, i.e., $G\left(\eta_{AF}\right)$ is integrated accordingly with respect to  bounds, as represented in (\ref{eq:21-1-1}).
\begin{algorithm*}[tbh]
\raggedright{}
\begin{equation}
\begin{aligned}F_{\rho}\left(\eta_{AF}\right)= & \intop_{0}^{\eta_{AF}}\left(1-\sum_{a=0}^{m_{\mathrm{SD}}-1}\sum_{b=0}^{a}\frac{\left(\frac{\eta_{AF}-y}{\eta_{AF}\left(1+y\right)}\right)^{b}\times\left(\tilde{\alpha}_{\mathrm{SD}}\eta_{AF}\right)^{a}}{b!(a-b)!}\times\frac{\Gamma\left(m_{\mathrm{RD}}+b\right)}{\left(\tilde{\alpha}_{\mathrm{RD}}+\left(\frac{\eta_{AF}-y}{1+y}\right)\tilde{\alpha}_{\mathrm{SD}}\right)^{(m_{\mathrm{RD}}+b)}}\times\frac{e^{(-\tilde{\alpha}_{\mathrm{SD}}\eta_{AF})}\times\tilde{\alpha}_{\mathrm{RD}}^{m_{\mathrm{RD}}}}{\Gamma(m_{\mathrm{RD}})}\right)\times f_{\gamma_{\mathrm{SR}}}(y)\,dy+\\
 & \intop_{\eta_{AF}}^{\infty}\Biggl(\frac{\gamma\left(m_{RD},\tilde{\alpha}_{\mathrm{RD}}\left(\frac{1+y}{y-\eta_{AF}}\right)\eta_{AF}\right)}{\Gamma(m_{\mathrm{RD}})}-\sum_{a=0}^{m_{\mathrm{SD}}-1}\sum_{b=0}^{a}\frac{\left(\frac{\eta_{AF}-y}{\eta_{AF}\left(1+y\right)}\right)^{b}\times\left(\tilde{\alpha}_{\mathrm{SD}}\eta_{AF}\right)^{a}}{b!(a-b)!}\times\\
 & \frac{\gamma\left(m_{\mathrm{RD}}+b,\tilde{\alpha}_{\mathrm{RD}}\left(\frac{1+y}{y-\eta_{AF}}\right)\eta_{AF}-\tilde{\alpha}_{\mathrm{SD}}\eta_{AF}\right)}{\left(\tilde{\alpha}_{\mathrm{RD}}+\left(\frac{\eta_{AF}-y}{1+y}\right)\tilde{\alpha}_{\mathrm{SD}}\right)^{(m_{\mathrm{RD}}+b)}}\times\frac{e^{\left(-\tilde{\alpha}_{\mathrm{SD}}\eta\right)}\times\tilde{\alpha}_{\mathrm{RD}}^{m_{\mathrm{RD}}}}{\Gamma(m_{\mathrm{RD}})}\Biggr)\times f_{\gamma_{\mathrm{SR}}}(y)\,dy,
\end{aligned}
\label{eq:21-1-1}
\end{equation}
\end{algorithm*}
Finally, (\ref{eq:21-1-1}) is manipulated, to get thus, the AF outage
expression, i.e., $F_{\rho}\left(\eta_{AF}\right)$ as depicted
in (\ref{eq:21}).

\begin{algorithm*}[tbh]
\raggedright{}
\begin{equation}
\begin{aligned}F_{\rho}\left(\eta_{AF}\right)= & 1-\frac{\tilde{\alpha}_{\mathrm{SR}}^{m_{\mathrm{SR}}}}{\Gamma(m_{\mathrm{SR}})\Gamma(m_{\mathrm{RD}})}\intop_{\eta_{AF}}^{\infty}y^{m_{\mathrm{SR}}-1}e^{-\tilde{\alpha}_{\mathrm{SR}}y}\times\Gamma\left(m_{\mathrm{RD}},\tilde{\alpha}_{\mathrm{RD}}\left(\frac{1+y}{y-\eta_{AF}}\right)\eta_{AF}\right)dy-\\
 & \frac{e^{(-\tilde{\alpha}_{\mathrm{SD}}\eta_{AF})}\times\tilde{\alpha}_{\mathrm{RD}}^{m_{\mathrm{RD}}}\times\tilde{\alpha}_{\mathrm{SR}}^{m_{\mathrm{SR}}}}{\Gamma(m_{\mathrm{RD}})\Gamma(m_{\mathrm{SR}})}\sum_{a=0}^{m_{\mathrm{SD}}-1}\sum_{b=0}^{a}\frac{\left(\tilde{\alpha}_{\mathrm{SD}}\eta_{AF}\right)^{a}}{b!(a-b)!}\times\\
 & \Biggl(\intop_{0}^{\infty}\frac{\Gamma\left(m_{\mathrm{RD}}+b\right)\left(\frac{\eta_{AF}-y}{\eta_{AF}\left(1+y\right)}\right)^{b}y^{m_{\mathrm{SR}}-1}e^{-\tilde{\alpha}_{\mathrm{SR}}y}}{\left(\tilde{\alpha}_{\mathrm{RD}}+\left(\frac{\eta_{AF}-y}{1+y}\right)\tilde{\alpha}_{\mathrm{SD}}\right)^{(m_{\mathrm{RD}}+b)}}dy-\intop_{\eta_{AF}}^{\infty}\frac{\left(\frac{\eta_{AF}-y}{\eta_{AF}\left(1+y\right)}\right)^{b}y^{m_{SR}-1}e^{-\tilde{\alpha}_{\mathrm{SR}}y}}{\left(\tilde{\alpha}_{\mathrm{RD}}+\left(\frac{\eta_{AF}-y}{1+y}\right)\tilde{\alpha}_{\mathrm{SD}}\right)^{(m_{\mathrm{RD}}+b)}}\times\\
 & \Gamma\left(m_{\mathrm{RD}}+b,\tilde{\alpha}_{\mathrm{RD}}\left(\frac{1+y}{y-\eta_{AF}}\right)\eta_{AF}-\tilde{\alpha}_{\mathrm{SD}}\eta_{AF}\right)dy\Biggr)
\end{aligned}
\label{eq:21}
\end{equation}
\end{algorithm*}
The integrals generally, do not generate a closed form expression,
thus, it can be evaluated numerically using matlab software.
\begin{itemize}
\item \textbf{Selective Decode and Forward}
\end{itemize}
In the following, we briefly introduce the FD SDF relay system outage
probability. In SC mode, if the relay correctly decodes the received
packet, and decides to re-transmit the re-encoded block through the
$\mathrm{R}\rightarrow\mathrm{D}$ link, both relay and direct links
are combined at the destination side. This mode's outage probability
is generally, denoted as the same form as (\ref{eq:sdfout}). However,
the threshold will be redefined accordingly as, $\eta_{\mathrm{SDF}}=2^{R\left(\frac{NL+\tau_{SDF}}{NL}\right)}-1$\footnote{The factor $\frac{NL+\tau_{\mathrm{SDF}}}{NL}$ means that the transmission
of $NL$ useful bits occupies $NL+\tau_{\mathrm{SDF}}$ channel uses. }, and $\mathit{P_{\mathrm{out}}^{\mathrm{XD}}}$ in $(\ref{eq:sdfout})$
will be denoted, $\mathit{P_{\mathrm{out}}^{\mathrm{S}\mathcal{\mathrm{R}}\mathrm{D}}}$,
which represents the outage probability of the combined signal, i.e.,
direct and relayed signals, at the destination side, and it can be
derived as follows, 

\begin{equation}
\mathit{P_{\mathrm{out}}^{\mathrm{S}\mathcal{\mathrm{R}}\mathrm{D}}}=\mathrm{Pr}\left(\frac{1}{NL+\tau_{SDF}}\sum_{i=0}^{NL-1}\log_{2}(1+\gamma_{i})<r\right).\label{eq:14-1}
\end{equation}
Hence, by referring to \cite{13,14}, the end-to-end SINR, can be
approximated to $\gamma_{i}\approx\gamma_{\mathrm{SD}}+\gamma_{\mathrm{RD}}$
. Thus, using (\ref{eq:approx_snr}) and (\ref{eq:linkSNRNC}), the
expression of $\mathit{P_{\mathrm{out}}^{\mathrm{S}\mathcal{\mathrm{R}}\mathrm{D}}}$
is approximated and given by, 

\begin{align*}
\mathit{P_{\mathrm{out}}^{\mathrm{S}\mathcal{\mathrm{R}}\mathrm{D}}}\approx & \mathrm{Pr}\left(\frac{NL}{NL+\tau_{SDF}}\log_{2}\left(1+\gamma_{\mathrm{SD}}+\gamma_{\mathrm{RD}}\right)<r\right).
\end{align*}
Therefore, $P_{\mathrm{out}}^{\mathrm{S}\mathcal{\mathrm{R}}\mathrm{D}}$
can be derived as: 
\begin{align}
\mathit{P_{\mathrm{out}}^{\mathrm{S}\mathcal{\mathrm{R}}\mathrm{D}}}= & \mathrm{Pr}\left(\gamma_{\mathrm{SD}}+\gamma_{\mathrm{RD}}<\eta_{\mathrm{SDF}}\right)\nonumber \\
= & \intop_{0}^{\eta_{\mathrm{SDF}}}\mathrm{Pr}\left(\gamma_{\mathrm{SD}}<\eta_{\mathrm{SDF}}-y\right)\times f_{\gamma_{\mathrm{RD}}}(y)\textrm{d}y,\label{eq:}
\end{align}

Hereafter, while developing the integral form, we got the expression
of $P_{out}^{\mathrm{S}\mathcal{\mathrm{R}}\mathrm{D}}$, as given
in \cite{14},

\begin{align}
\mathit{P_{out}^{\mathrm{S}\mathcal{\mathrm{R}}\mathrm{D}}}= & \frac{\gamma(m_{\mathrm{RD}},\tilde{\alpha}_{\mathrm{RD}}\eta_{\mathrm{SDF}})}{\Gamma(m_{\mathrm{RD}})}-\sum_{k=0}^{m_{\mathrm{SD}}-1}\left(\tilde{\alpha}_{\mathrm{SD}}\eta_{SDF}\right)^{k}\times\frac{\left(\tilde{\alpha}_{\mathrm{RD}}\eta_{SDF}\right)^{m_{\mathrm{RD}}}\times e^{-\left(\tilde{\alpha}_{\mathrm{RD}}\eta_{SDF}\right)}}{\Gamma(m_{\mathrm{RD}}+k+1)}\nonumber \\
\times & _{1}F_{1}\left(k+1;m_{\mathrm{RD}}+k+1;\tilde{\alpha}_{\mathrm{RD}}-\tilde{\alpha}_{\mathrm{SD}}\right).\label{eq:srd}
\end{align}
where $_{1}F_{1}\left(k+1;m_{\mathrm{RD}}+k+1;\tilde{\alpha}_{\mathrm{RD}}-\tilde{\alpha}_{\mathrm{SD}}\right)$
denotes the Kummer\textquoteright s confluent hypergeometric function.

Finally, by substituting (\ref{eq:SD_cdf}), (\ref{eq:SR_SD}) and
(\ref{eq:srd}) into (\ref{eq:sdfout}), we get the closed form expression
of SDF outage probability.  

\section{Numerical Results }\label{sec:Results-and-discussion}
In this section, the theoretical findings derived in section \ref{sec:Amplify-and-Forward-Outage-Proba},
are numerically verified and confirmed using Monte-carlo simulations.
The variation of outage probability for various transmission schemes
investigated in section \ref{sec:System-Model}, is represented. Moreover,
for an exhaustive comparison, simulations include the direct transmission
mode with no relay cooperation. To assess SC-SDF performances in term
of super block length, we consider two cases: 1) the case where the
super-block length is very large compared with the relay processing
delay, $N\times L\gg \tau_{SDF}$ , 2) and the case of short super-block where $L=3$\footnote{The super-block length is set while respecting a low latency requirement less than 1$\,\textrm{ms}.$ }. In this section, the first case is denoted SC-SDF while the second case is denoted SC-SDF3.
For all simulations, we consider a packet length of $N=20$ symbols, and a relay processing delay of $\tau_{AF}=1$ symbol for symbol-by-symbol transmissions and $\tau_{SDF}=500\,\textrm{\ensuremath{\mu}s}$ \footnote{According to a 3rd Generation Partnership Project 3GPP study on latency
reduction techniques for LTE, the latency induced for encoding and
decoding processing is proportional to the block size, and it represents
3 times the block size \cite{21}. Therefore, $\tau_{SDF}=3\times N\times T_{\textrm{S}}$,
with $T_{\textrm{S}}$ represents the duration of one symbol. In this work, we consider $T_{\textrm{S}}=8,33\,\textrm{\ensuremath{\mu}s}$ which represents a typical symbol duration in Millimeter-wave (mmWave)
bands with subcarrier spacing of 120$\,\textrm{KHZ}$ \cite{22}.} for block-by-block transmissions.  $P_{\mathrm{S}}=P_{\mathrm{R}}=5\,\textrm{dB}$ and $m_{\textrm{SD}}=m_{\textrm{SR}}=m_{\textrm{RD}}=2$.

\begin{figure}[tbh]
\begin{centering}
\includegraphics[scale=0.6]{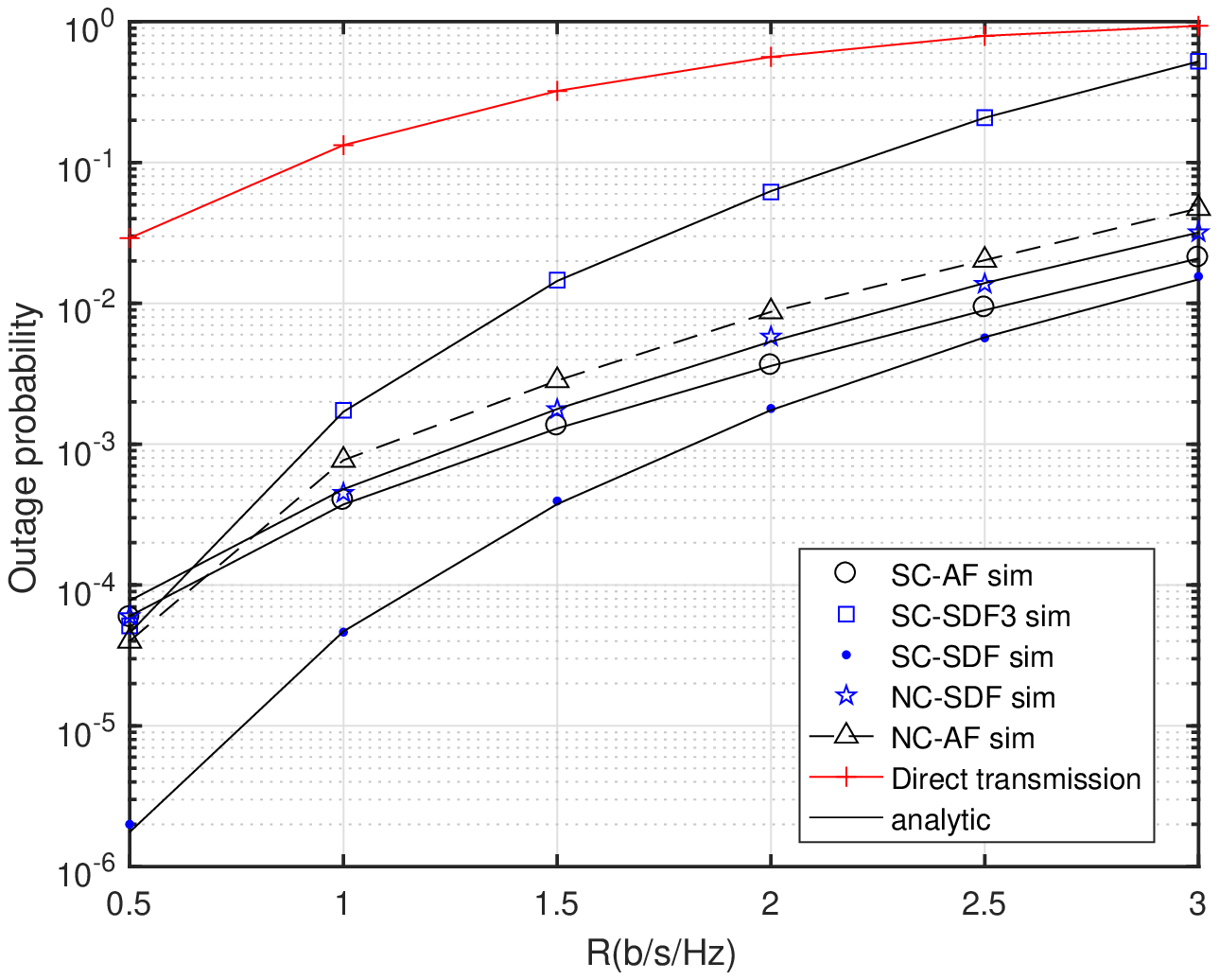}

\caption{\label{fig:Outage-probability-versusCASE1bitratelowSD}Outage probability
versus $R$ , for $\sigma_{\mathrm{SR}}^{2}=\sigma_{\mathrm{RD}}^{2}=20\,\mathrm{dB}$,
$\sigma_{\mathrm{SD}}^{2}=0\,\mathrm{dB}$ and $\sigma_{\mathrm{RR}}^{2}=0\mathrm{\,dB}$.}

\includegraphics[scale=0.6]{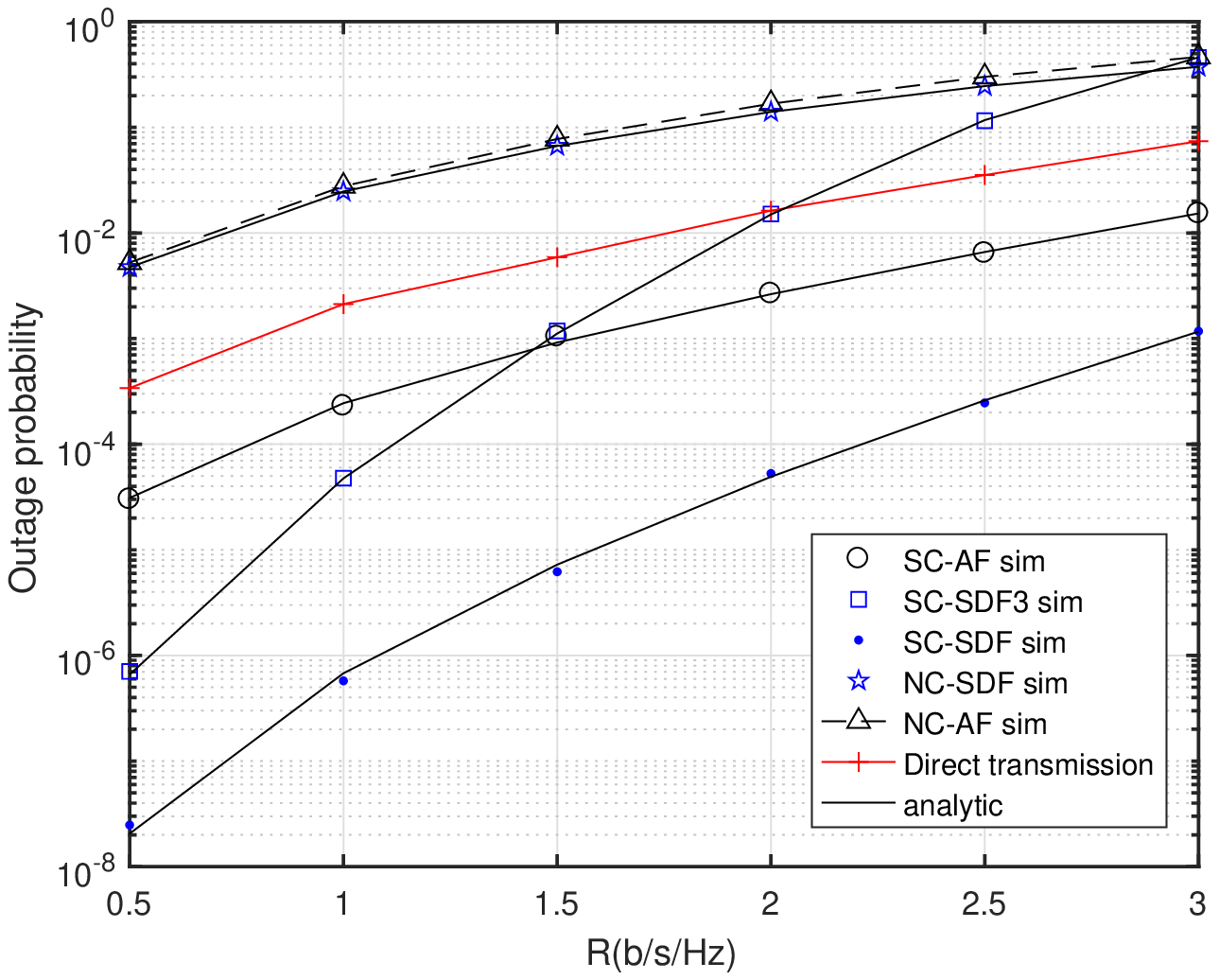}
\par\end{centering}
\caption{\label{fig:Outage-probability-versusCASE1bitratehighSD-1}Outage probability
versus $R$ , for $\sigma_{\mathrm{SR}}^{2}=\sigma_{\mathrm{RD}}^{2}=20\,\mathrm{dB}$,
$\sigma_{\mathrm{SD}}^{2}=10\,\mathrm{dB}$ and $\sigma_{\mathrm{RR}}^{2}=0\mathrm{\,dB}$.}
\end{figure}

First, we compare the studied relaying schemes performances in term
of the spectral efficiency level. For that purpose, in Fig.\ref{fig:Outage-probability-versusCASE1bitratelowSD}
and Fig.\ref{fig:Outage-probability-versusCASE1bitratehighSD-1},
we plot the outage probability versus the transmission bit rate R.
First, we notice that the simulation results confirm the accuracy
of the analytical expressions, obtained in section \ref{sec:Amplify-and-Forward-Outage-Proba}.
In both figures, we note that, for transmissions that support very
long super-block, i.e., $\frac{NL}{NL+\tau_{SDF}}\rightarrow1$, the
SC-SDF offers the best performances. However, for transmissions with low latency requirements less than $1$ ms,
the super-block length must be less than $L=3$. Thus, using SC-SDF
is not anymore the obvious choice. Thereafter, in term of the low latency purpose, we see that, in Fig. \ref{fig:Outage-probability-versusCASE1bitratelowSD}, when the direct link gain is very low compared to the first and second hop gains, the low processing delay scheme, SC-AF scheme, offers the best performance. However, as the direct link gain increases, we start to notice that SC-SDF3 becomes more desirable for low transmission rate, while the SC-AF still the best choice for high transmission rate. This is mainly due to the SDF rate penalty of $\frac{NL}{NL+\tau_{SDF}}$ that impacts banefully the spectral efficiency, whenever the super block size decreases.

\begin{figure}[tbh]
\begin{centering}
\includegraphics[scale=0.6]{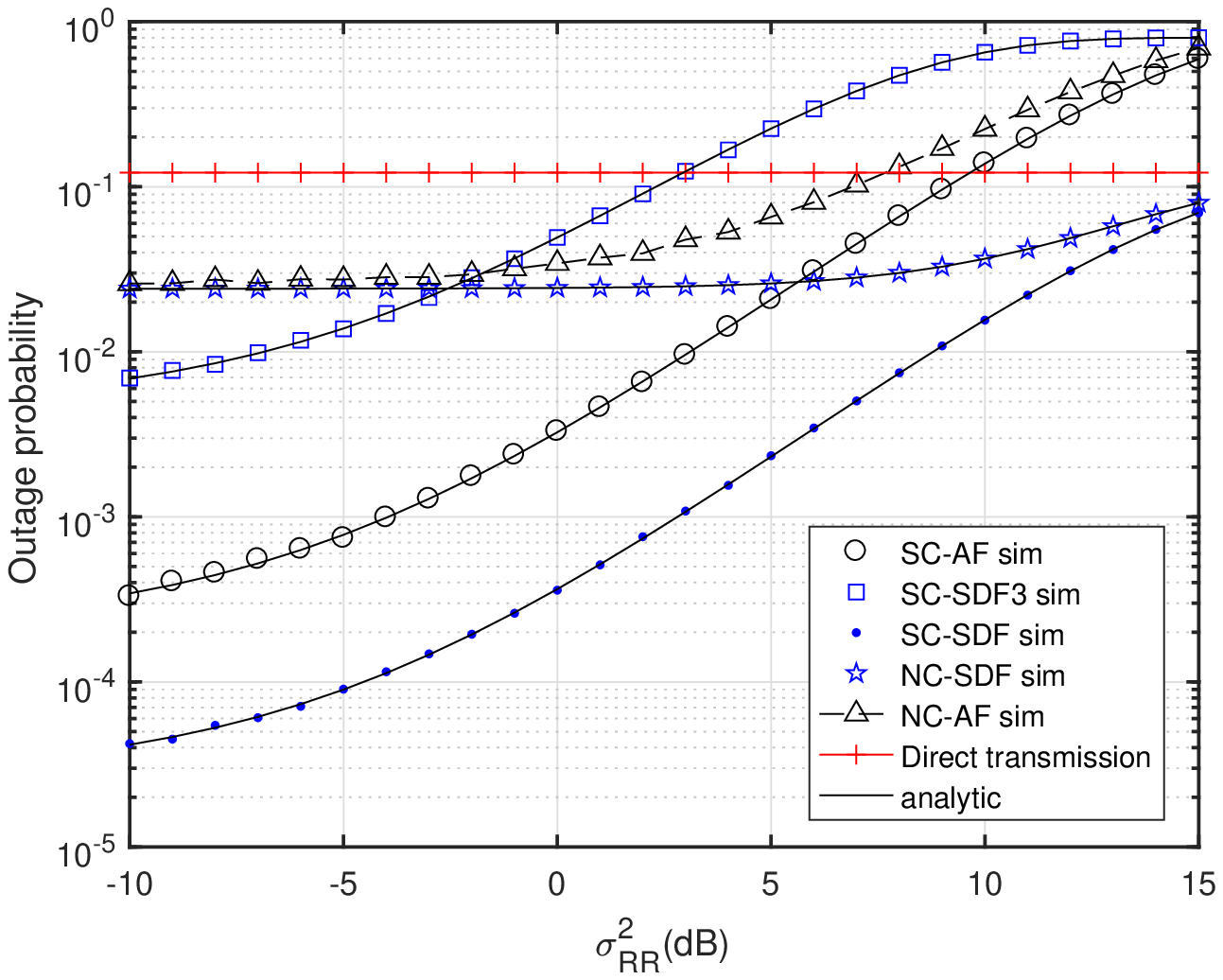}

\caption{\label{fig:Outage-probability-RSIhighSRRDlowSD}Outage probability
versus $\sigma_{\mathrm{RR}}^{2}$ , for $R=2$bps/Hz, $\sigma_{\mathrm{SR}}^{2}=\sigma_{\mathrm{RD}}^{2}=20\,\mathrm{dB}$
and $\sigma_{\mathrm{SD}}^{2}=5\,\mathrm{dB}$ }

\includegraphics[scale=0.6]{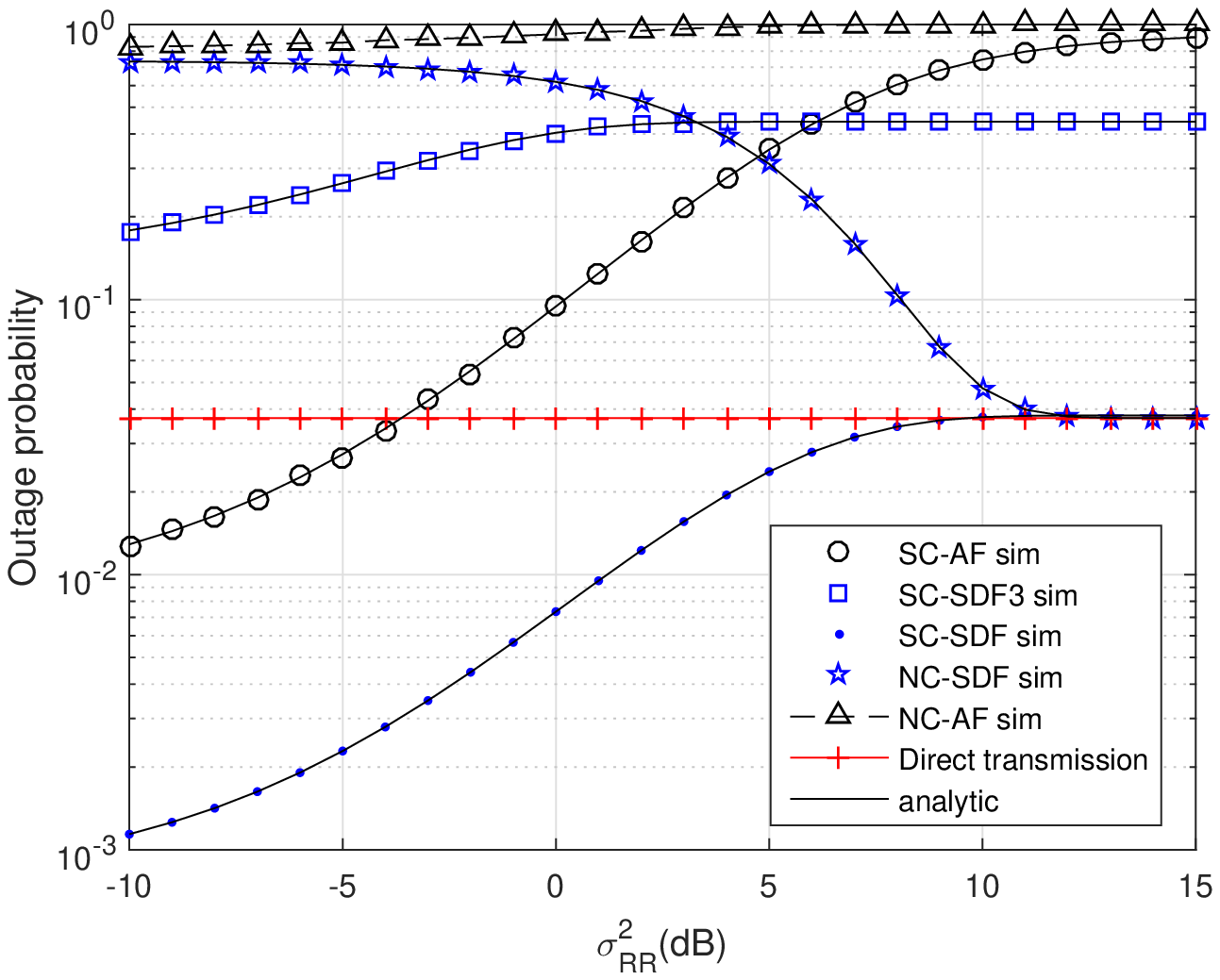}
\par\end{centering}
\caption{\label{fig:Outage-probability-RSIapproxSRRDSD}Outage probability
versus $\sigma_{\mathrm{RR}}^{2}$ , for $R=2$bps/Hz, $\sigma_{\mathrm{SR}}^{2}=\sigma_{\mathrm{RD}}^{2}=10\,\mathrm{dB}$
and $\sigma_{\mathrm{SD}}^{2}=8\,\mathrm{dB}$ .}
\end{figure}

Hereafter, to point out the impact of the RSI level on performances,
Fig. \ref{fig:Outage-probability-RSIhighSRRDlowSD} and Fig. \ref{fig:Outage-probability-RSIapproxSRRDSD}
illustrate the outage probabilities as function of $\sigma_{RR}^{2}$.
In one hand, we see clearly that SC-SDF still provides the best performance.
However, this scheme can not be practically adopted for low latency
transmissions. In the other hand, we notice that there are three transmission
schemes that outperform each other depending on the RSI level at the
relay and can be practically adopted for low latency transmissions:
SC-AF, NC-SDF, and direct transmission. In fact, SC-AF seems to be
the most suitable scheme for low latency transmissions with low RSI
at the relay, i.e., $\sigma_{RR}^{2}\leq5$ dB. However, for moderate
and high RSI, i.e., $\sigma_{RR}^{2}>5\,\textrm{dB}$, we can either
use NC-SDF if the direct link is not strong enough (Fig. \ref{fig:Outage-probability-RSIhighSRRDlowSD})
or just switch-off the relay if the direct link gain is as good as
the relay link (Fig. \ref{fig:Outage-probability-RSIapproxSRRDSD}).
In fact, in Fig. \ref{fig:Outage-probability-RSIapproxSRRDSD}, we
see that the direct transmission clearly outperforms NC-SDF scheme.
This is due to the fact that, in NC mode, the destination will try
to decode the strongest received signal while the remaining signal
will be considered as interference. Accordingly, at low RSI, where
the relay can correctly decode and forward the re-encoded block, the
destination will receive a useful signal as strong as the interfering
signal, which dramatically deteriorates the system performances. As
the RSI gain increases, i.e., $\sigma_{RR}^{2}>5\,\textrm{dB}$, the
relay fails to correctly decode the received packet. Therefore, the
only received signal at destination is the direct link signal. That
is clearly seen in Fig. \ref{fig:Outage-probability-RSIapproxSRRDSD}
where the NC-SDF curve improves, as the $\sigma_{RR}^{2}$ increases,
to be similar to the direct transmission curve.

\begin{figure}[tbh]
\begin{centering}
\includegraphics[scale=0.6]{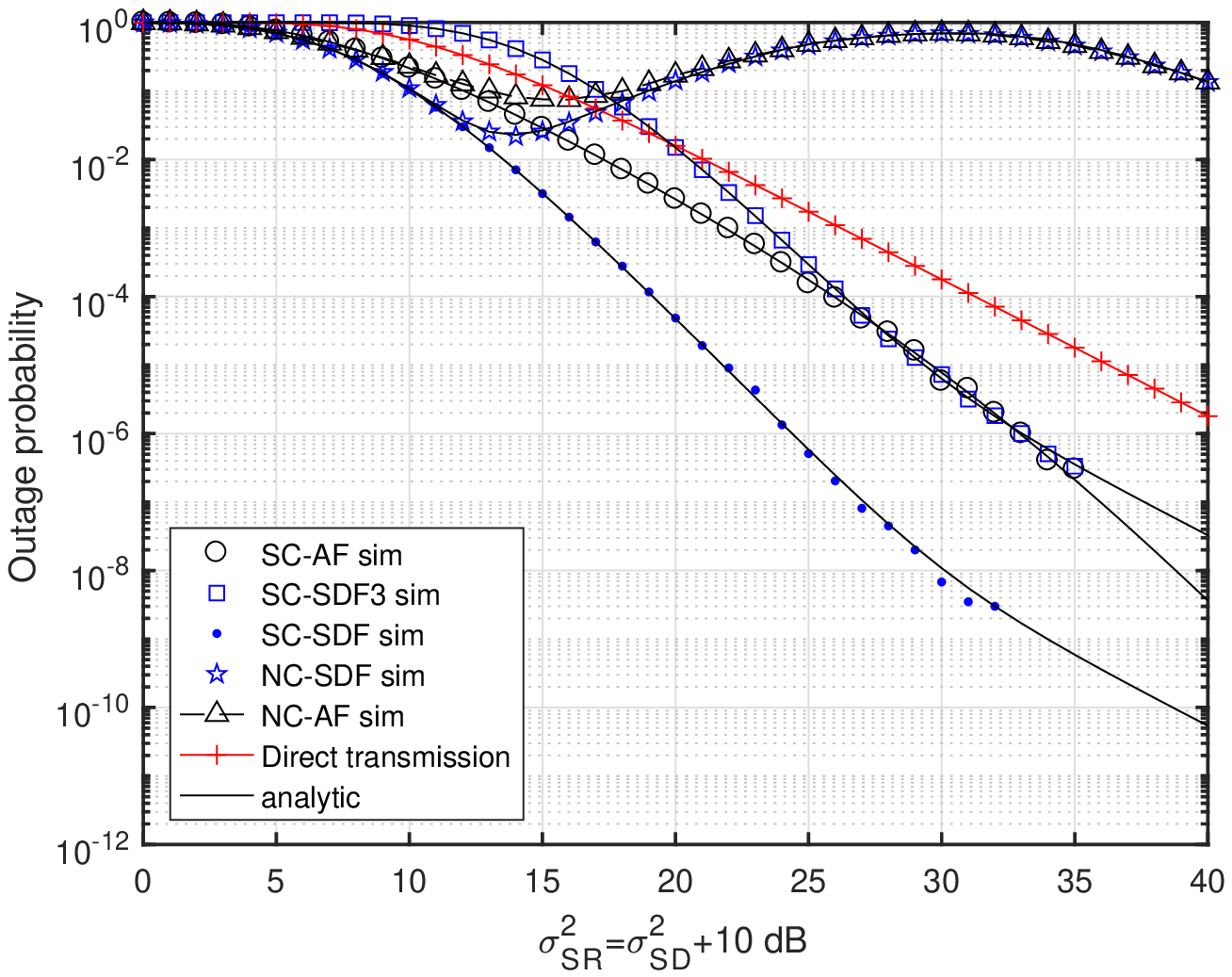}

\caption{\label{fig:Outage-probability-versusCASE1highRD}Outage probability
versus $\sigma_{\mathrm{SR}}^{2}$ , for $R=2$bps/Hz, $\sigma_{\mathrm{RD}}^{2}=20\,\mathrm{dB}$
and $\sigma_{\mathrm{RR}}^{2}=0\mathrm{\,dB}$.}

\includegraphics[scale=0.6]{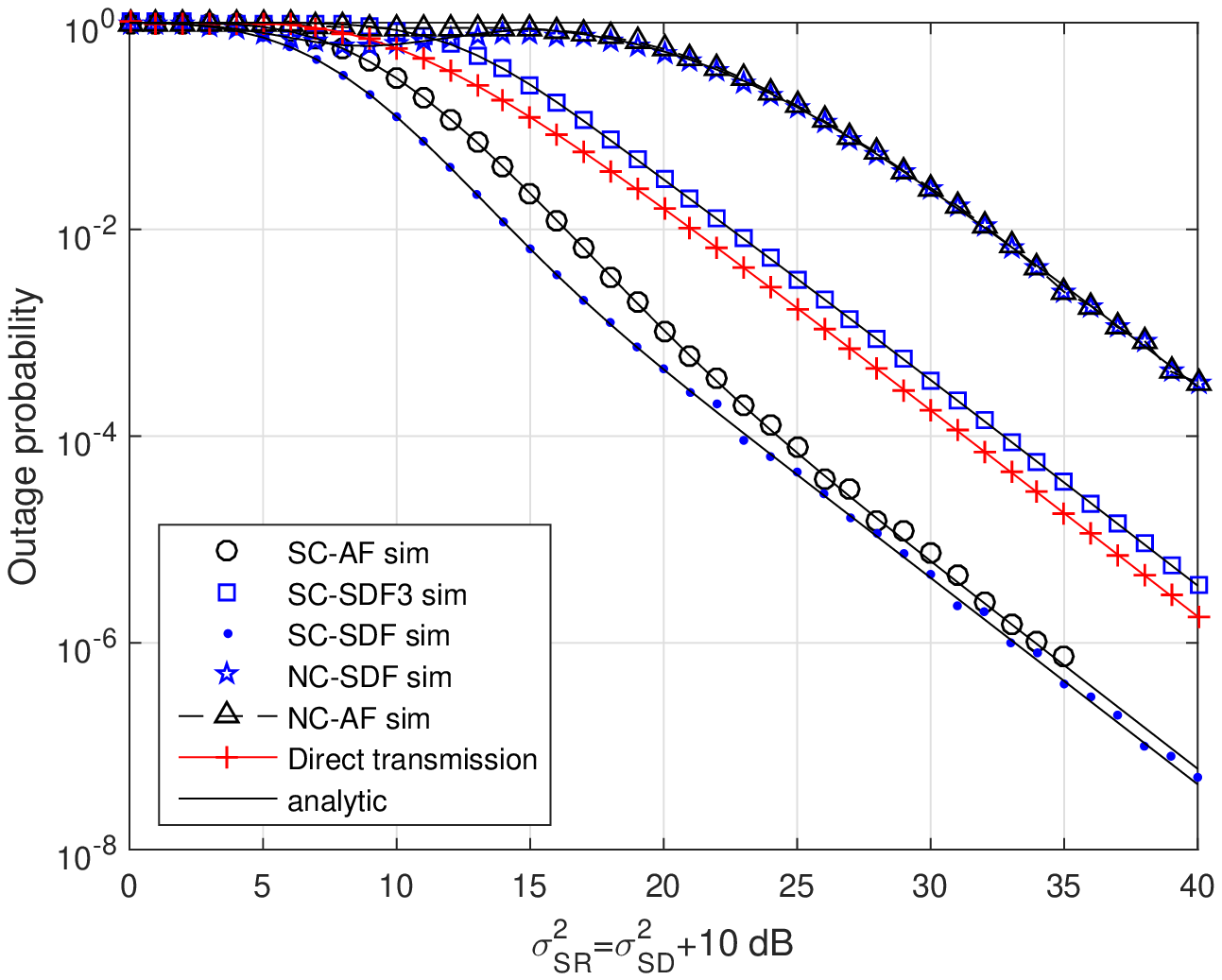}
\par\end{centering}
\caption{\label{fig:Outage-probability-versusCASE1lowRD}Outage probability
versus $\sigma_{\mathrm{SR}}^{2}$, for $R=2$bps/Hz, $\sigma_{\mathrm{RD}}^{2}=5\,\mathrm{dB}$
and $\sigma_{\mathrm{RR}}^{2}=0\mathrm{\,dB}$.}

\end{figure}

Now, we consider the scenario where the $\mathrm{S}\rightarrow\mathrm{R}$
link is much better than the $\mathrm{S}\rightarrow\mathrm{D}$ link,
i.e., $\sigma_{\mathrm{SR}}^{2}=\sigma_{\mathrm{SD}}^{2}+10\,\mathrm{dB}$.
Fig.\ref{fig:Outage-probability-versusCASE1highRD} and Fig.\ref{fig:Outage-probability-versusCASE1lowRD}
plot the outage probability versus a range of $\mathrm{S}\rightarrow\mathrm{R}$
link gains. We see clearly that, for moderate $\mathrm{R}\rightarrow\mathrm{D}$
link quality, i.e., $\sigma_{\mathrm{RD}}^{2}=5\,\mathrm{dB}$, SC-AF
scheme offers better outage performance than all other studied schemes.
On the other hand, as the $\mathrm{R}\rightarrow\mathrm{D}$ link
variance increases, i.e., $\sigma_{\mathrm{RD}}^{2}=20\mathrm{\,dB}$,
we notice that the performance of NC modes are enhanced for low $\mathrm{S}\rightarrow\mathrm{D}$
link gain, i.e., $\sigma_{\mathrm{SD}}^{2}<5\,\mathrm{dB}$, where
the direct link harmful impact becomes negligible. Indeed, a performance
gap between the NC-SDF and the SC-SDF3 can be specifically noticed,
when the $\mathrm{S}\rightarrow\mathrm{D}$ link gain becomes lower,
at above, $\sigma_{\mathrm{SD}}^{2}<6\,\mathrm{dB}$. Therefore, in
this case, if considering the SDF protocol, it is rather better to
choose the NC mode than the SC. While with strong $\mathrm{S}\rightarrow\mathrm{D}$
link, the SC modes are more suitable for transmission scenarios, mainly
due to the additional spatial diversity. Moreover, the SC-AF presents
higher performances for higher $\mathrm{S}\rightarrow\mathrm{R}$
link gains, which is obvious as a result.

\section{Conclusion}\label{sec:Conclusion}

Two basic cooperative relaying schemes, i.e. AF and SDF, were studied
over a Nakagami-m fading channel, where one FD relay assisted the
communication between a source node and a destination node. For an
exhaustive comparison, we adopted both relaying protocols over two
different transmission mode, i.e., the non-combining mode and the
signals combining mode. Simulations results, proved that the SDF block-based
transmission scheme is no longer practical to adopt for FD SC mode,
where direct and relay links are combined at the receiver side, especially,
with the more latency induced due to the complexity of encoding and
decoding algorithms. Still the AF scheme represents better choice
in term of outage performance and latency. On the other hand, SDF
relaying scheme is more suitable for non combining transmission mode.

\nocite{*}
\clearpage

\end{document}